\documentclass[twocolumn]{aastex63}

\usepackage{comment}
\usepackage{lipsum}
\usepackage{threeparttable}
\usepackage{xcolor}
\usepackage{colortbl}
\usepackage{color}
\usepackage{array}
\usepackage{multirow}
\usepackage{bigstrut}
\usepackage{amsfonts}
\usepackage{amsmath}
\usepackage{amssymb}
\usepackage[pangram]{blindtext}
\usepackage{tabu}
\usepackage{threeparttable}
\usepackage{mathrsfs}
\usepackage{textcomp}
\DeclareUnicodeCharacter{2212}{-}
\DeclareUnicodeCharacter{00A0}{~}

\definecolor{LightCyan}{rgb}{0.88,1,1}
\newcolumntype{a}{>{\columncolor{LightCyan}}c}

\newcommand{\swift}{\textit{Swift}}

\shorttitle{SED modeling of M81*}
\shortauthors{Tomar and Gupta}
\graphicspath{{./}{figures/}}

\begin{document}

\title{X-Ray Flares in the Long-term Light Curve of Low-luminosity Active Galactic Nucleus M81*}

\correspondingauthor{Gunjan Tomar}
\email{gunjan@rri.res.in}

\author[0000-0003-4992-6827]{Gunjan Tomar}
\affiliation{Astronomy \& Astrophysics group, Raman Research Institute, C. V. Raman Avenue, 5th Cross Road, Sadashivanagar, Bengaluru 560080, Karnataka, India}

\author[0000-0002-1188-7503]{Nayantara Gupta}
\affiliation{Astronomy \& Astrophysics group, Raman Research Institute, C. V. Raman Avenue, 5th Cross Road, Sadashivanagar, Bengaluru 560080, Karnataka, India}

\begin{abstract}
Most active galactic nuclei at the center of the nearby galaxies have supermassive black holes accreting at sub-Eddington rates through hot accretion flows or radiatively inefficient accretion flows, which efficiently produce jets. The association of radio and X-ray flares with the knot ejection from M81* inspires us to model its multiwavelength spectral energy distribution (SED) during these flares to constrain the physical parameters of the jet. Moreover, we construct a long-term light curve in X-rays to identify the flares in the available data and constrain the jet parameters during those periods. The jet activity may vary on short as well as long timescales, which may produce flares in different frequency bands. The SEDs from radio to X-ray during the quiescent as well as flaring states are found to be satisfactorily explained by synchrotron emission of relativistic electrons from a single zone. The variation in the values of the jet parameters during the different states is shown and compared with high-synchrotron peaked blazars.
\end{abstract}

\section{Introduction} 
\label{sec:intro}
Most giant galaxies are known to host supermassive black holes (SMBHs) at their centers. In our local universe, most of these SMBHs are underfed due to low, sub-Eddington accretion rates and are classified as low-luminosity active galactic nuclei (LLAGNs). Unlike typical active galactic nuclei (AGNs), which have a thin disk, LLAGNs have radiatively inefficient accretion flow (RIAF) at their inner radii and a truncated thin disk at the outer radii. Observations and theoretical studies suggest that these accretion flows are quite efficient at producing bipolar outflows and jets. Therefore, it has been suggested that in most of these LLAGNs, emission comes from three components: a jet, RIAF, and an outer thin disk. Their relative contribution to the spectral energy distributions (SEDs) of LLAGNs is not yet known. The low luminosity of these sources makes them incapable of sustaining structural features like a broad-line region (BLR) and dusty torus; ergo, they do not follow the unification model of AGNs.
\par
The SMBHs spend most of their lifetime as LLAGNs, but these are not as well studied as their brighter counterparts due to their faintness. Due to its proximity, M81* is relatively bright and has been one of the most extensively observed LLAGNs. Thus, it allows us to explore the underlying physical mechanisms of the otherwise less-known LLAGNs.
\par
M81* is located at a distance of 3.6 Mpc (\citealt{freedman}) with a SMBH of estimated mass 7$\times10^7$ M$_{\odot}$, at the center of the massive spiral galaxy NGC 3031. The bolometric luminosity of M81* is 9.3$\times10^{40}$ ergs/s, implying an Eddington ratio of M81* is 9.3$\times10^{40}$ ergs/s implying an Eddington ratio of 1.1$\times10^{-5}L_{Edd}$, where $L_{Edd}$ is the Eddington luminosity. The low Eddington ratio, the absence of a thin disk at the inner radius (\citealt{young}), and its proximity make it a prototype LLAGN to study the emission mechanisms and temporal features of LLAGNs, which yet remain unexplored.
\par
Jet-dominated spectral models for M81* have previously been used to explain the data from radio to X-ray frequencies (\citealt{markoff}), with both synchrotron and synchrotron self-Compton (SSC) as the proposed dominant mechanisms of X-ray emission, providing a good fit to the observed data. It is important to understand whether the jet activity of M81* is similar to the jet activity of the general population of high-luminosity AGNs, which have been extensively studied in the recent past with multi-wavelength data from various telescopes.
\par
A uniquely large radio flare (\citealt{poole}) was seen to be followed by a discrete knot ejection in M81*, suggesting a similar jet-production mechanism to luminous AGNs (\citealt{king}). A X-ray flare and a radio rebrightening were found to be associated with the ejected knot. Their result has a significant impact on the understanding of the jet activity of M81*, as this indicates for the first time the presence of radial motion in the jet of a low-luminosity AGN.
\par
We constrain the physical parameters of the jet during the emission of the discrete knot in radio frequencies by constructing and modeling the multiwavelength SEDs. Moreover, we identify several X-ray flares in the long-term light curve of M81* (2005 April 21–2014 May 24) constructed using Swift data. The methods followed to analyze the X-ray, UV- optical and gamma-ray data are given in Section \ref{sec:methods}, and the methods used to identify the flares and quiescent state are discussed in Section  \ref{sec:lc}. We discuss the construction of the SEDs during the different states in Section \ref{sec:sedc} and, subsequently, the modeling of these SEDs is discussed in Section \ref{sec:model}.
\par 
Our results are mentioned in Section \ref{sec:results} and discussed in Section \ref{sec:discussions}. Interestingly, we find that all the SEDs covering radio to X-ray frequencies can be explained by synchrotron emission from the jet of M81*, similar to high-synchrotron peaked blazars (HSPs) like Mrk 421 (\citealt{sinha2015}). 

\section{Multiwavelength observations and Data Analysis}
\label{sec:methods}
\subsection{X-ray} 
\label{sec:xray}
The \swift~telescope has monitored M81* several times between 2005 April 21 to 2021 May 24. The X-ray data were taken from the X-Ray Telescope (XRT) instrument onboard \swift~, in both photon counting (PC) and windowed timing (WT) modes. The data reduction was performed using the standard data pipeline package \texttt{(XRTPIPELINE v0.13.5)} in order to produce cleaned event files. For the observations in WT mode, the source events were extracted using a circular aperture of radius 118\arcsec, and the background events were extracted from an annular region with an inner and outer radius of 200\arcsec.6 and 271\arcsec.4, respectively, chosen to be symmetrically placed about 100 pixels. The count rate for each of these observations was below the pile-up limit ($\sim$ 100 counts s$^{-1}$). For the observations in PC mode, the source events were extracted from a circular region of radius 118 \arcsec centered on the source. The background events were extracted from an annular region of inner radius and outer radius of 141 \arcsec and 213 \arcsec, respectively (\citealt{king}). The observations with the count rate above the pile-up limit ($\sim$ 0.5 counts s$^{-1}$) were checked for pile-up effect by modeling the XRT-PSF with Kings function following the standard procedure. The source event files for the observations with pile-up were then extracted by excluding the events within the circle of radius up to which pile-up was significant (varying between 2\arcsec and 10 \arcsec). The spectra were obtained from the corresponding event files using the \texttt{XSELECT v2.4g} software; we created the response files using the task \texttt{xrtmkarf} and then combined them with the source and background spectra using \texttt{grppha}. Due to low count rates, we binned the spectra with a minimum count of 1 per bin using \texttt{grppha}. We used \texttt{XSPEC} to fit each spectrum using Cash statistics. We fit each spectrum with a simple absorbed power law \texttt{(tbabs*po)} with corresponding Galactic hydrogen column density (N$_H$) along the direction of M81* fixed to 5 $\times$ 10$^{20}$ cm$^{-2}$ (\citealt{king}). The power law is defined as $A(E)=KE^{-\Gamma_x}$, where the photon index $\Gamma_x$ and the normalization K were allowed to vary. The flux obtained in soft X-ray band (0.5-2.0 keV) and hard X-ray band (2.0-10.0 keV) at each epoch is shown in Figures \ref{lc}(a) and \ref{lc}(b), respectively. 
\par
Observation taken by another X-ray instrument, \textit{NuSTAR} during the period of the detected flares was also analyzed for completeness.
\par
An observation was taken by \textit{NuSTAR} on 2015 May 18 during Flare B. The data were processed using \texttt{NUSTARDAS (v1.8.0)} along with \texttt{CALDB v20190627}. The cleaned event files with an exposure of $\sim$ 20.9 ks were obtained using \texttt{nupipeline}. We extracted the source spectrum from a circular region of radius 100$''$ centered on the source for both Focal Plane Modules A and B (FPMA and FPMB). The background spectrum was extracted from a region of an equivalent radius away from the source. We carried out the spectral fitting over an energy range of 3.0-60.0 keV for FPMA and FPMB simultaneously using XSPEC with the same baseline model along with three Gaussian absorption lines with centroid energies fixed at 6.4, 6.68, and 6.96 keV (\citealt{page}).

\subsection{UV-Optical} 
M81* was monitored by \textit{Swift}'s Ultra-violet Optical Telescope (UVOT) in some or all of six filters, V (5468 \AA), B (4392 \AA), U (3465 \AA), W1 (2600 \AA), M2 (2246 \AA) and W2 (1928 \AA) simultaneously with multiple XRT observations. We reduced the data using the standard procedures, by extracting source counts from a circular region of 5$''$ radius centered on the source, while background counts were extracted from an annulus region centered on the source with inner and outer radii of 27$''$.5 and 35$''$, respectively. We used the tool \texttt{uvotsource} to derive the magnitudes, which were then corrected for Galactic extinction using the extinction values obtained with python module \texttt{extinction} and the reddening law with $R_v =$ 3.1 from \citet{fitz99}. The corrected observed magnitudes were further converted into fluxes using the zero-point correction flux (\citealt{breeveld}). The \textit{Swift-UVOT} light curve is shown in Figures \ref{lc}(c) and \ref{lc}(d). The host galaxy flux (\citealt{pian}) was further subtracted for U, W1, M2 and W2 filters.
\subsection{Gamma-Ray}
We analyzed the data collected by \textit{Fermi}'s Large Area Telescope over a period of $\sim$ 13.5 yr ranging from 2008 August 8 to 2022 February 28 with \texttt{fermitools v2.0.0}, \texttt{fermipy v1.0.0}, and Pass 8 event processed data. A strong gamma-ray emitter, starburst galaxy M82, is located at 0.62$^\circ$ away from M81*. Since the point spread function (PSF) of \textit{Fermi-LAT} at lower energies is larger than the separation, contamination from M82 to gamma-ray flux cannot be ruled out (\citealt{lenain}). Therefore, we selected the events in the 10-300 GeV energy range, where the PSF is smaller than the separation. The events are selected in a 15$^\circ$ × 15$^\circ$ region of interest (ROI) centered on the position of M81*. The data were binned spatially with a scale of 0.1$^\circ$ per pixel and eight logarithmically spaced bins per energy decade. We then manually added a point source at the center of the ROI, as M81* does not belong to the 4FGL catalog. The source was modeled with a power law spectrum.
\par
We only selected the \texttt{Source} class events (\texttt{evclass = 128} and \texttt{evtype = 3}) with the recommended filter expression (\texttt{DATA\_QUAL$>$0} \&\& \texttt{LAT\_CONFIG $==$ 1}). Also, a maximum zenith angle cut of 90$^{\circ}$ was applied to reduce the contamination from secondary gamma-rays from the Earth limb.
\par
We included the standard diffuse templates, “\texttt{gll\_iem\_v07}" and “\texttt{iso\_P8R3\_SOURCE\_V2\_v1}”, available from the Fermi Science Support Center \footnote{\url{https://fermi.gsfc.nasa.gov/ssc/data/access/lat/BackgroundModels.html}}, to model the Galactic diffuse emission and the residual background and isotropic extragalactic emission, respectively.
\par
A binned maximum-likelihood analysis was performed by taking into account all the sources included in the updated fourth source catalog (4FGL-DR2) and lying up to 5$^{\circ}$ outside the ROI in order to obtain the spectral parameters and the significance of detection of the source.
\par
Automatic optimization of the ROI was performed using function \texttt{optimize} within the package to ensure that all the parameters were close to their global likelihood maxima. To look for any additional sources in our model which were not included in the 4FGL catalog, we used \texttt{find\_sources()} with a power law model with index 2, \texttt{sqrt\_ts\_threshold = 5.0}, and \texttt{min\_seperation = 0.5}. Additional sources, when detected with TS$>$25 were included during the LAT analysis. 
\\
The normalization of all the sources within a radius of 5$^{\circ}$ from the ROI and the isotropic and Galactic diffuse emission templates were left to vary. We derived an upper limit on gamma-ray flux of M81* as 2.58 $\times$ 10$^{-13}$ ergs cm$^{-2}$ s$^{-1}$ over 10-300 GeV energy range.

\section{Long-term Light Curve}
\label{sec:lc}
 We construct a long-term light curve by plotting the absorption-corrected soft (0.5 - 2 keV) and hard (2.0-10.0 keV) X-ray flux and optical and UV flux with the date of observations in Figure \ref{lc}.
\subsection{Identication of Flares}
\label{sec:baye}
\begin{figure*}
\centering
\includegraphics[width=\textwidth]{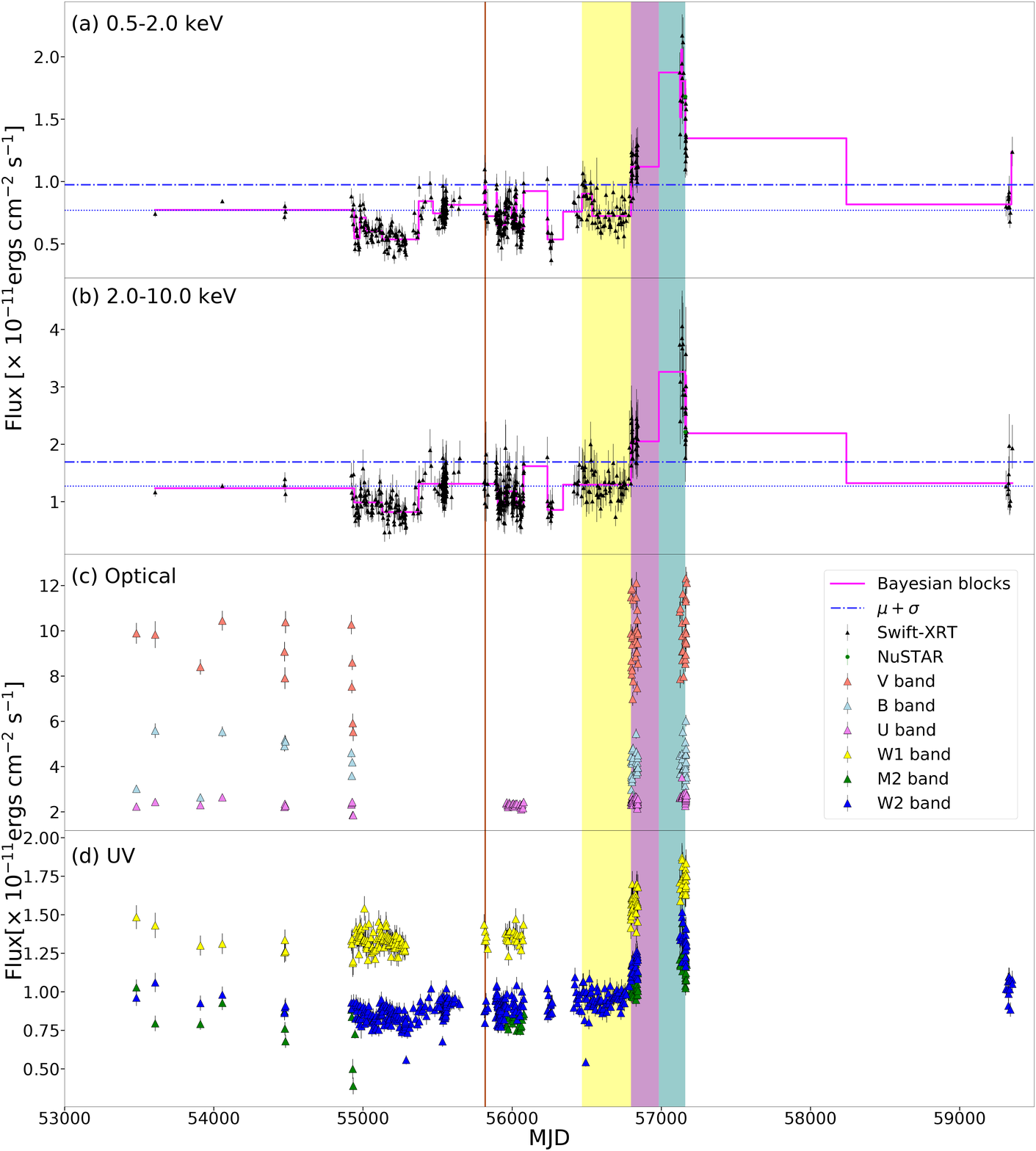}                                                       
\caption{The lightcurve in (a) soft X-ray energies (0.5-2.0 keV) and (b) hard X-ray energies (2.0-10.0 keV), and the light curve in the (c) optical and (d) UV. The yellow region represents the quiescent state (MJD 56467.7 - MJD 56796.7). The brown line (MJD 55814.83 - MJD 55823.35), and the purple (MJD 56796.74 - MJD 56983.37) and teal (MJD 56983.37 - MJD 57160.25) shaded regions represent Flare A, Flare B, and Flare C, respectively.}
\label{lc}
\end{figure*}

\begin{figure*}
\centering
\includegraphics[width=\textwidth]{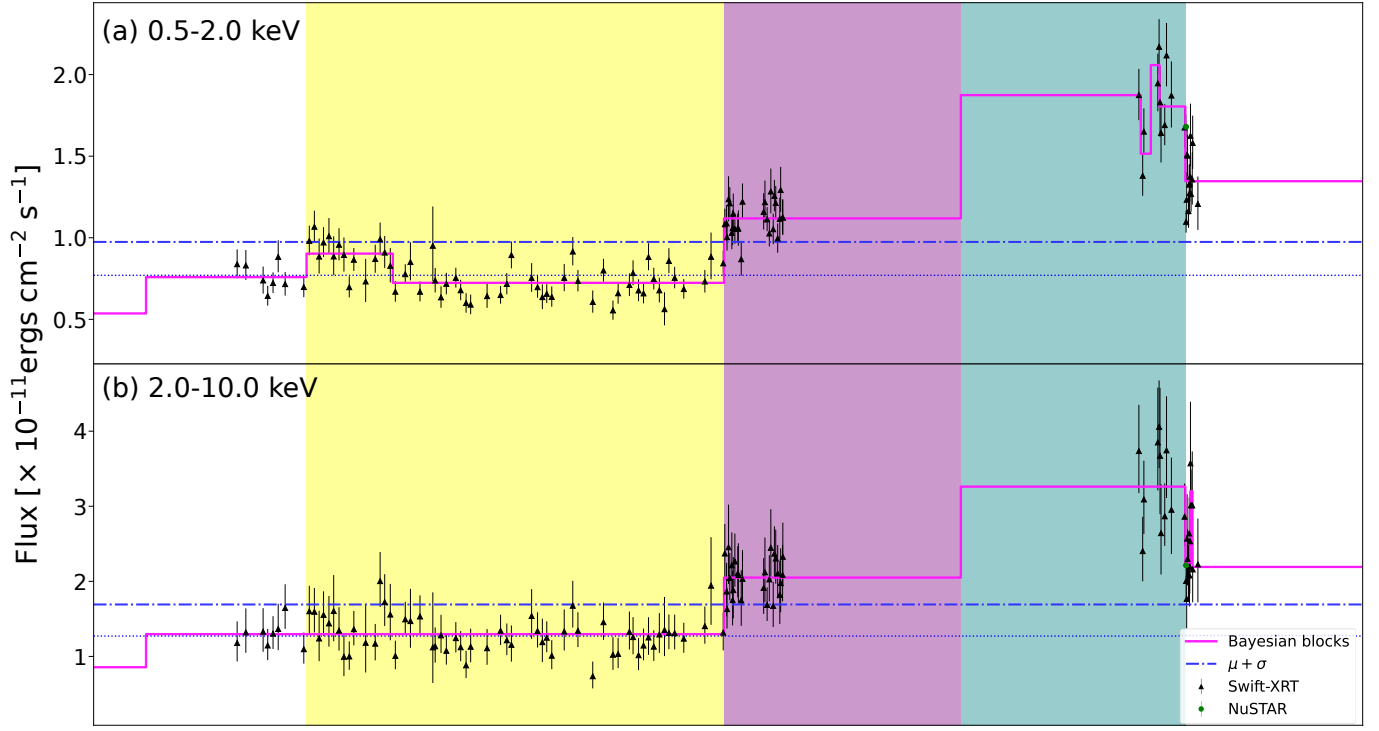}                                                       
\caption{Same as Figures \ref{lc}(a) and (b), zoomed in to the region of Flare B and Flare C to show the change in flux.}
\label{lcf}
\end{figure*}

For the identification of statistically significant X-ray flares in the light curve, we use the Bayesian Block algorithm (\citealt{Scargle}). The algorithm finds the optimally spaced time intervals by taking into account the statistical fluctuations from the flux measurement errors. We use the \texttt{astropy} implementation of the Bayesian Block algorithm with a false-positive rate of 0.01 with the option of ``measures" in the fitness function. The change in flux states as obtained by the Bayesian Block algorithm is represented by a magenta line in Figure \ref{lc}. The height of each block represents the statistical mean of all the flux measurements within that block. We define the statistical mean of all the flux measurements as the base (or quiescent) flux. If the mean flux value in a block exceeds the base value by a factor of n$\sigma$, we consider it as a flare, where n is an integer and $\sigma$ is the standard deviation. By this definition, we identify two flares, Flare B and Flare C, seen in both soft and hard X-ray energies, as shown in Figures \ref{lc} and \ref{lcf}. We also consider Flare A in our study since it was identified as an X-ray flare preceding a radio rebrightening that followed the largest radio flare observed for M81* by \cite{king}, even though it exceeds the base flux value by only 0.8$\sigma$ (shown by the red line in Figure \ref{lc}). The corresponding flux in the higher energy range is close to the baseline flux which is consistent with no flare seen in this band by \cite{king}, as well.

\subsection{Identification of Quiescent State Period} 
We select the time period of MJD 56467.7- MJD 56796.7 for the quiescent state. The flux in this time period is close to the mean average flux over the entire time period of the light curve. Also, the response matrix files of the observations of this period are the same, thus eliminating the need to sum different response files.

\section{Multiwavelength Spectral energy Distributions}

\subsection{Construction of Spectral Energy Distributions for Different States}
\label{sec:sedc}
 In Figure \ref{lc-nt}, we zoom in to the time period over which rebrightening of M81* was observed in the radio band in 2011 (\citealt{king}). Though an increase in flux in the soft X-ray band is seen, it is not found to be statistically significant as discussed in Section \ref{sec:baye}. However, to estimate the jet parameters and the electron distribution and their evolution with time during knot ejection, we consider this period as Flare A. The green dashed lines represent the four epochs over which radio observations were recorded by \cite{king}. We refer to them as Epochs 1, 2, 3, and 4, respectively, in our paper. Since Epoch 1 lies within the flaring period of Flare A, we include this to construct a quasi-simultaneous SED for Flare A. 
Similarly, simultaneous and quasi-simultaneous SEDs are constructed for Epochs 2, 3, and 4.
\par
To construct a SED for Epoch 2, we have taken the \swift~ observation recorded on the same day (2011 September 21). For Epoch 3 (2011 September 27), due to a lack of observation by \swift~, we construct SED with only radio measurements. For Epoch 4 (2011 October 4), the \swift~ observation recorded on 2011 October 3 has been used for constructing the SED.
\par
We constrain the emission in the gamma-ray band with the upper limit on gamma-ray flux from M81*. The compilation of multiwavelength data for the different epochs and flares is shown in Figure \ref{sed}, which includes both archival data and the data analyzed in the present work (shown in color). 
\subsection{Modeling of Spectral Energy Distributions}
\label{sec:model}
 We model the compiled broadband SEDs in different states using the time-dependent code \texttt{GAMERA\footnote{\url{http://libgamera.github.io/GAMERA/docs/main_page.html}}} to investigate the evolution of the broadband spectrum, in particular, the X-ray spectrum. We consider a homogeneous and spherical emission region (blob) of radius R having the magnetic field B and moving down along the jet with a bulk Lorentz factor $\Gamma$. This region contains relativistic electrons having a power law distribution of particles and emitting radiation through synchrotron and SSC processes.
Using this code, we calculate the particle spectrum $N= N(E,t)$ under the continuous injection of the particles described by $Q(E,t)$ and energy loss rate $b=b(E,t)$.
The code solves the transport equation,

\begin{equation}
    \frac{\partial N}{\partial t} = Q(E,t) - \frac{\partial (bN)}{\partial E} - \frac{N}{t_{esc}}
\end{equation}

 where $t_{esc}=\frac{R}{c}$ is the timescale over which the leptons escape from the emission region.  
The code subsequently calculates the synchrotron and SSC emission, which are Doppler boosted by a factor of $\delta^4$ in the observer's frame due to relativistic beaming, where $\delta = [\Gamma(1-\beta cos\theta)]^{-1}$ is the Doppler factor, $\Gamma$ is the bulk Lorentz factor of the emission region or the jet frame, $\beta$ is the intrinsic speed of the emitting plasma, and $\theta$ is the viewing angle of the jet with respect to the line of sight of the observer.
For modeling, we consider a quiescent state as the time at which the system attains a steady state, i.e, $N(E,t)$ no longer evolves with time.
For Epochs 2, 3, and 4, the time over which the electron spectrum evolves is constrained by the number of days between each epoch. 
For each flare, we allow $N(E,t)$ to evolve over the time period of the respective flare obtained with the Bayesian block method. For all the SEDs, the Lorentz bulk factor has been kept at a fixed value of 3.7 (\citealt{doi}). 
We vary the parameters that define the injected electron distribution e.g., the spectral index, the minimum Lorentz factor ($\Gamma_{min}$), the maximum Lorentz factor ($\Gamma_{max}$) and the normalization factor (A), and also the parameters of the emission region e.g., the radius of the emission region (R) and the magnetic field (B) in that region to fit the data. 
Here, the primed quantities denote the parameters in the comoving frame of the emission region in the jet, and the unprimed quantities are the values of the parameters measured in the observer's frame. The energy density of the injected electrons is $U_{e,inj}'$ in the comoving frame, and the jet power in the injected electrons, P$_{e,inj}$, is calculated with the following expression:

\begin{equation}
P_{e,inj} = \pi R^2 \Gamma^2 c U_{e,inj}' = \frac{3\Gamma^2c}{4R}\int_{E_{min}}^{E_{max}} EQ(E)dE
\end{equation}

\section{Results}
\label{sec:results}
We identify three flares (Flare A, Flare B, and Flare C) in the soft X-ray band as shown in Figure \ref{lc}. A flare is observed in the high X-ray band for both Flare B and Flare C but not for Flare A. In Figures \ref{lc}(c) and (d), we show the flux in optical and UV bands as obtained by \swift~-UVOT. The flux in UV and optical bands is also found to be increased during Flare B and Flare C. This flux is corrected for Galactic extinction and host-galaxy contribution. Therefore, the rise in flux in these bands during Flare B and Flare C points toward variability in M81*. 
\par
Interestingly, the X-ray flux shows a significant variation at different time intervals (Figure \ref{sed}). Using time-dependent SED modeling, the evolution of the jet parameters has been studied in each state.
\par
By modeling the SED during the quiescent state (2013 June 24 - 2014 May 19), we find that the radio to X-ray emission can be explained by one-zone synchrotron emission, which extends up to 10$^{20}$ Hz, while the SSC emission is well below the upper limits that we have obtained by Fermi-LAT in the gamma-ray energy band from 10 to 300 GeV. The system attains a steady state after 200 days. It is noted that the hard spectrum of X-rays in the quiescent state requires a hard injected electron spectrum with spectral index p$=$2.42 (Figure \ref{sed-fit}(a)). The size of the emitting region is 0.029 pc, and the system is found to be particle dominated with $U_B'$/$U_e' \approx$ 0.017, two orders of magnitude away from the equipartition of energy in the jet. The jet parameter values used for our SED modeling are shown in Table \ref{table_par}. The total jet power during the quiescent state is 8.84$\times10^{41}$ ergs s$^{-1}$, while it shows little variation during the other epochs and flares. 
\par
Subsequently, we perform the SED modeling for all the other states, as shown in Figure \ref{sed-fit}(a)-(g). Synchrotron emission can explain the radio to X-ray data for all the epochs and flares. All the SEDs simulated from our model are presented together in Figure \ref{sed-fit}(h) to show the variation in spectral shape and flux during the different states of M81*.
\par
During Flare A, the X-ray spectrum hardens, which is explained by the hardening in the injected electron distribution. 
\par
\begin{figure*}
\centering
\includegraphics[scale=0.12]{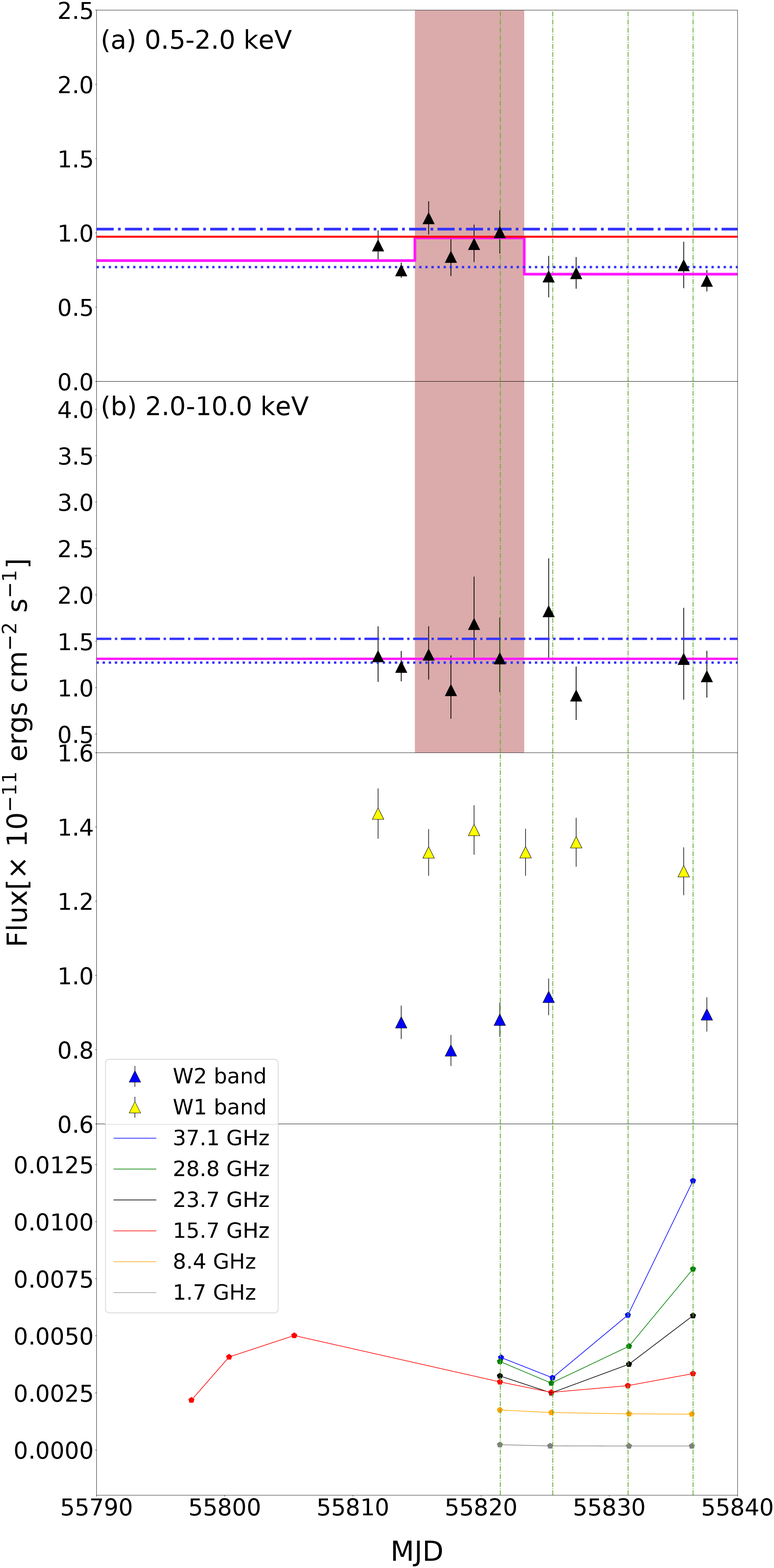}                                                       
\caption{The lightcurve during the time period considered at the time of knot ejection (\citealt{king}). The measurements in radio are taken from the same work. The black data points are \swift~ measurements. The magenta line in Panels (a) and (b), represent the Bayesian blocks, as in Figure \ref{lc}. The blue dotted line is the mean flux over the entire period of 16 yr that we have considered in our study. The red line in panel (a) is the mean flux plus 0.8$\sigma$, which represents the peak flux of the X-ray flare associated with the knot ejection, while the blue dotted-dashed line represents 1$\sigma$ over mean flux. The vertical dotted-dashed lines represent the four epochs: 2011 September 17, September 21, September 27, and October 4, in the same order. For Flare A, we consider the radio data taken at Epoch 1 (2011 September 17) as it lies within the flare period for a simultaneous radio to X-ray SED. We also construct the SEDs for Epochs 2, 3, and 4, when quasi-simultaneous single-epoch X-ray measurements are available.}
\label{lc-nt}
\end{figure*}
\begin{figure*}
\centering
\includegraphics[width=\textwidth]{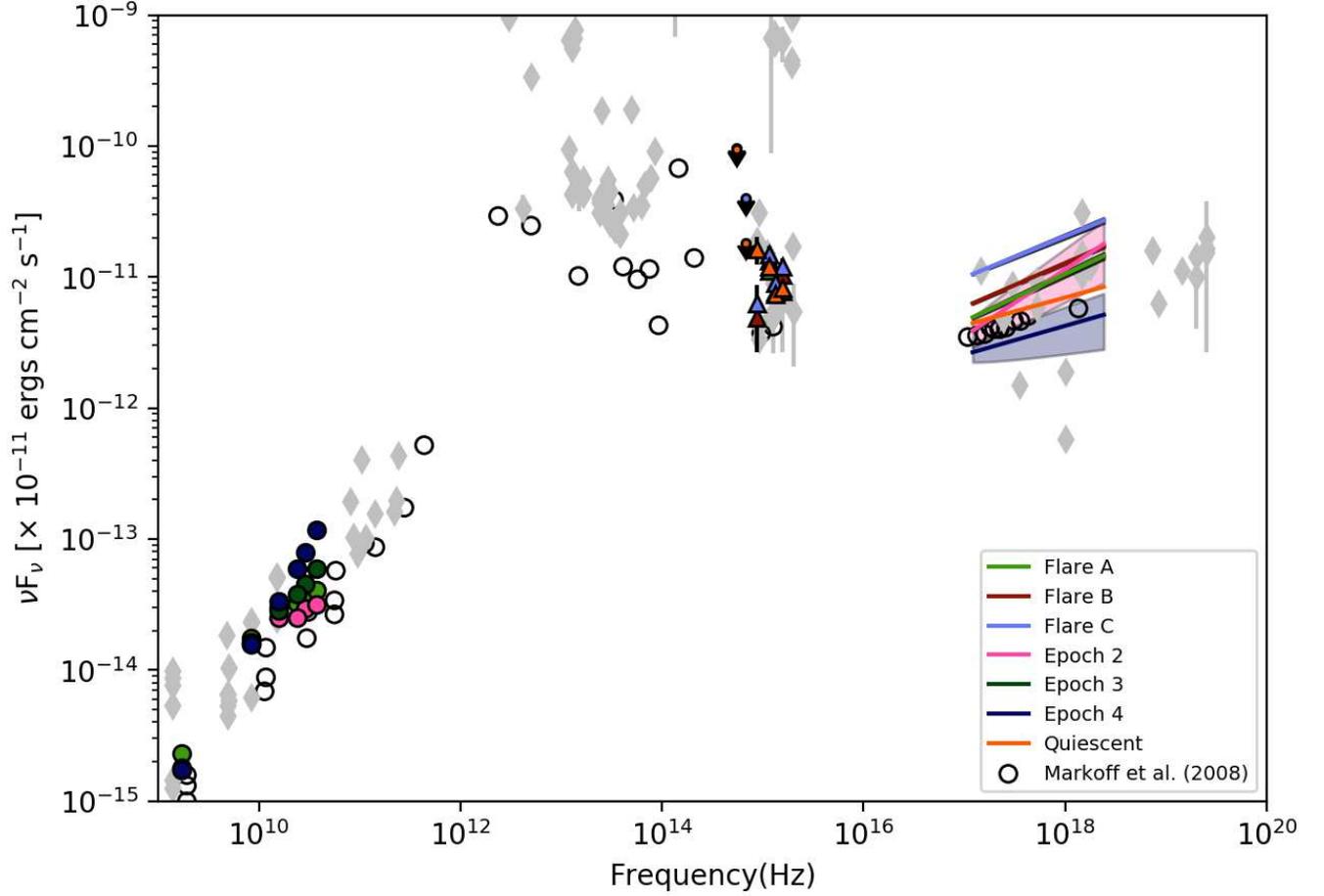}                                                     
\caption{The multi-wavelength data of M81* at different epochs and flaring periods are shown in color. In the X-ray band, the power law component of the X-ray spectrum obtained in these different states has been shown, with the shaded region representing the errors at 90\% confidence level (2.706$\sigma$). The simultaneous multi-wavelength data from \cite{markoff} at a quiescent period are also shown by open circles for reference. The gray diamonds are the archival data obtained from NASA/IPAC Extragalactic Database (NED), shown here as a secondary constraint for modeling at the frequencies where the simultaneous/quasi-simultaneous data are not available.}
\label{sed}
\end{figure*}
\begin{figure*}
    \centering
    \includegraphics[scale=0.28]{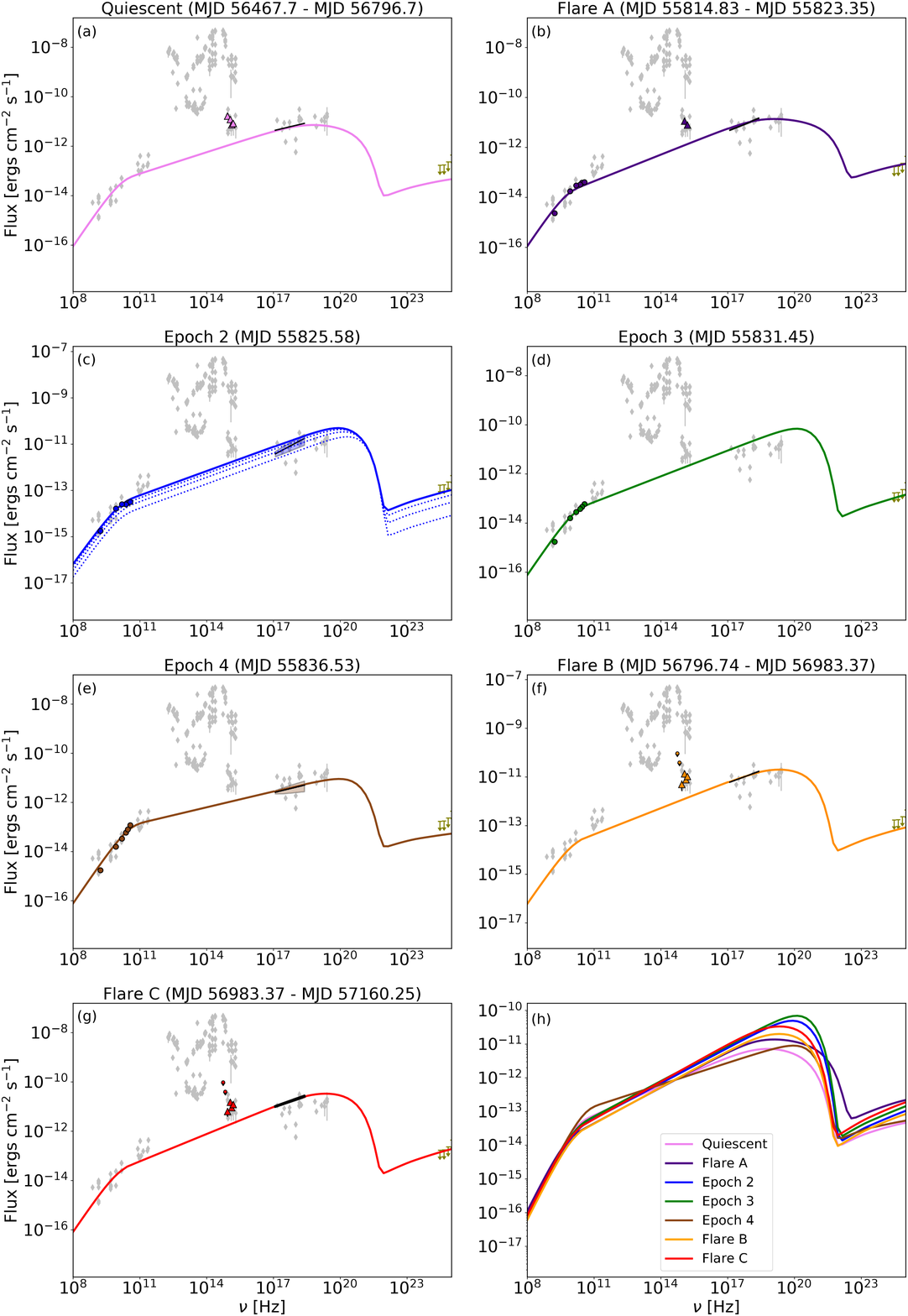}
    \caption{One-zone SED modeling of M81* during different states. Panel (c): the dotted curves in the modeling represent the evolution of SED from the initial time of injected electrons to the final age of the system, which explains the emission at this epoch. Panels (a)-(g): X-ray data for different states are shown in black. The upper limits obtained using Fermi-LAT data are shown in olive to constrain the SSC component and thus, the size of the emission region. Panel (h): Simulated SEDs  for all the states and epochs plotted together to show their variations in spectral shape and flux.}
    \label{sed-fit}
\end{figure*}

After 4 days, at Epoch 2, the radio spectrum has softened (as also seen in Figure \ref{lc-nt}), and the X-ray spectrum with similar flux has hardened. This is explained by change in injected electron energy density. 
\par
Six days later, on 2011 September 27 at Epoch 3, the radio spectrum becomes harder. Due to the lack of X-ray data at this epoch, we have fitted the radio spectrum with synchrotron emission using a parameter set similar to Epoch 2. It is found that, again, the injected electron density is increased to fit the radio data at this epoch.
\par
Eight days later, at Epoch 4, the X-ray spectrum is found to be softer than that at Epoch 2. It is interesting to note that, while the X-ray flux at this epoch is lower than that of the quiescent state, the radio flux becomes the maximum at this epoch (see Figure \ref{lc-nt}). The spectral index of the injected electron distribution required to explain the emission is the maximum at this epoch. Further, the injected electron density at this epoch is maximum. 
\par
Flare B (2014 May 19-2014 November 22) and Flare C (2014 November 22 - 2015 May 18) show flaring in both soft and hard X-ray bands. 


The values of the radius of the emission region (R) during Flare B and Flare C are the same as that during the quiescent state, and the magnetic field has been kept fixed at 0.003 Gauss, while the injected electron spectrum becomes harder.


\par
Epoch 4 requires the highest injected electron density among all the states as shown in Table \ref{table_par}. At all the epochs and states, the power in electrons has the highest contribution to the total jet luminosity. 
\par
The \swift~UV-Optical data points can be explained as multi-blackbody thermal emission from the disc (\citealt{markoff}, \citealt{lucchini}). 
\par
The variation of parameter values in the different states obtained from the X-ray data analysis (top two panels) and SED modeling is shown in Figure \ref{par}. 

\section{Discussions}
\label{sec:discussions}
Our time-dependent modeling of SEDs shows that the radio to X-ray emission observed from M81* can be well explained by synchrotron emission from the jet in all the states (quiescent to high X-ray flux states) and the epochs of radio knot emission. 
\subsection{Comparison with High-synchrotron Peaked Blazars}
The synchrotron emission peaks at 8.2$\times$10$^{18}$ Hz in the quiescent state of M81*, which is similar to extreme HSPs where the radio to X-ray data is explained with the synchrotron emission of relativistic electrons, with the peak frequencies higher than 10$^{17}$ Hz (\citealt{costamante01}, \citealt{costamante18}). For HSPs, eg. 1ES 1959+650 (\citealt{chandra21}), 1ES 1218+304 (\citealt{sahakyan20}), and Mrk 501 (\citealt{kk19}), it has been noted that the synchrotron peak frequencies, $\nu_{syn}$, can reach frequencies higher than 10$^{17}$ Hz, at least in the flaring states, showing extreme high-synchrotron-peaked behavior. We observe a similar behavior during the X-ray flares of M81*. The synchrotron peak shifts to higher frequencies with the increase in the amplitude of the X-ray flux, with the maximum value of $\nu_{peak}$ reaching up to  3.32$\times10^{19}$ Hz during Flare C. 

\subsection{Photon Index during Flares}
A harder-when-brighter behavior is commonly seen during X-ray flares of blazars, which manifests itself as a decrease in photon index with increasing flux (\citealt{xue06}). Figures \ref{par}(a) and (b), suggest a similar behavior in M81*. Our modeling shows that the spectral index of the injected electron distribution must be decreased with increasing X-ray flux (Figure \ref{par}(c)).
\par
\subsection{Magnetic Field during Flares}
The magnetic field required to fit the SEDs of M81* during all states is of the order of milliGauss, which is two orders of magnitude lower than the magnetic field required to fit the quiescent states of other LLAGNs detected in gamma-rays (see \citealt{gunjan}). However, in those cases, the X-ray data have been fitted with SSC emission except for M87, the synchrotron emission covers the X-ray data.
The magnetic field required to explain the synchrotron emission from HSP Mrk 501 during its high activity state in X-rays is of the order of 0.1 Gauss (\citealt{magic_mrk501_2020}) in the single zone leptonic model, which is similar to the magnetic field used in the modeling of several LLAGNs (\citealt{gunjan}) in the quiescent state. Ten extreme high-frequency peaked BL Lacs \citep{2020_tev}, which include 1ES 2037+521, TXS 0210+515, BZB 080+3455, TXS 0637-128, and six other sources were modeled with a single-zone conical-jet model. The magnetic field required to explain their X-ray data with synchrotron emission of leptons varies in the range of 0.02 to 0.25 G, which is comparable to Mrk 501 and higher than that of M81*.

\subsection{Role of Doppler Boosting in Flaring}
In the present work, the value of the Doppler factor 4.1 has been adapted from \citet{doi}, which is much lower than the typical value of the Doppler factor of the order of 10 observed in blazars. In other LLAGNs, the values of the Doppler factor are found to be 1.6 for NGC 315, 1 for NGC 4261, 2.3 for NGC 1275, and 2.3 for M87 (\citealt{gunjan}), which are comparable to that of M81*.

\par
Mrk 501 is at a distance of 140 Mpc, and M81* is only at a distance of 3.4 Mpc from us. If we compare their X-ray fluxes, during the quiescent state, the X-ray flux of Mrk 501 is a few times $10^{-10}$ erg cm$^{-2}$ sec$^{-1}$ (see Figure 1 of \cite{magic_mrk501_2020} and Figure 1 of \citealt{mrk501_ahnen17}), whereas the X-ray flux M81* is about 10$^{-11}$ erg cm$^{-2}$ sec$^{-1}$. During the flaring state, the X-ray and gamma-ray flux becomes a factor of 2 or higher than the flux in the quiescent state in the case of Mrk 501, but in the case of M81* a factor of 2 variations in X-ray flux is observed only during Flare C (which is the only X-ray flare with a significance of more than 3$\sigma$). During flares, the X-ray and gamma-ray flux of the blazars may become much higher compared to the same during their quiescent state as happened during the intense TeV flare of 1ES 1959+650 in 2016 (\citealt{1es_magic20}). In Figure \ref{comparison}, we compare the SED of 1ES 1959+650 with that of M81* during quiescent and flaring states. The unusually high gamma-ray flux has been explained with a small emission region and high value of Doppler factor in the range of 40 to 60.

\par
It is important to mention here that the value of the Doppler factor plays an important role in producing higher flux and higher variability in flux in blazars.
Due to high values of the Doppler factor, a higher flux of photons is expected from blazars in all energy bands compared to LLAGNs as the flux is Doppler boosted by a factor of $\delta^4$ in the observer's frame. Moreover, as a result of high values of the Doppler factor, the increase in flux during flaring states with respect to the quiescent state is higher compared to LLAGNs.

\subsection{Other Parameters during Flares}
In the long-term X-ray light curve of M81*, the variation in the X-ray flux is within a factor of 2 of the quiescent state flux, as a result, we do not report any large variation in injected power of electrons during Flare A, Flare B, and Flare C when the magnetic field is nearly constant. During Epochs 2, 3, and 4, the rise in the radio flux requires considerably higher power in injected electrons compared to the quiescent state. 

\par

\begin{figure*}
    \centering
    \includegraphics[scale=0.56]{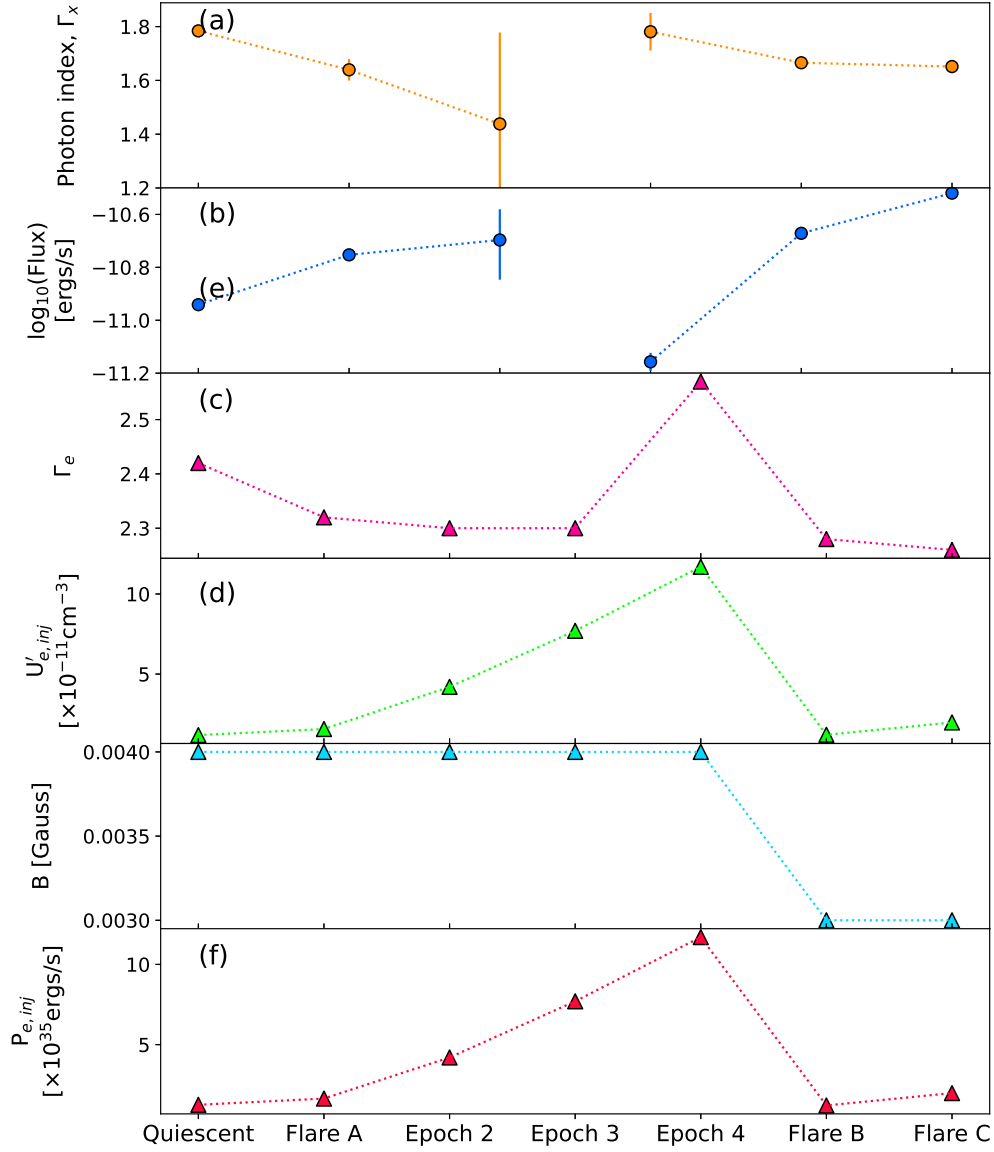}
    \caption{Time evolution of observables, (a) X-ray photon index, $\Gamma_x$, and (b) soft X-ray flux, and the parameters obtained from modeling, (c) injected electron spectral index, (d) energy density of injected electrons, (e) the magnetic field in the emission region, and (f) the jet power in injected electrons. Note that we do not have any quasi-simultaneous X-ray data for Epoch 3, so there is a break in the connected dot plot in the top two panels}.
    \label{par}
\end{figure*}

\begin{figure*}
    \centering
    \includegraphics[scale=0.56]{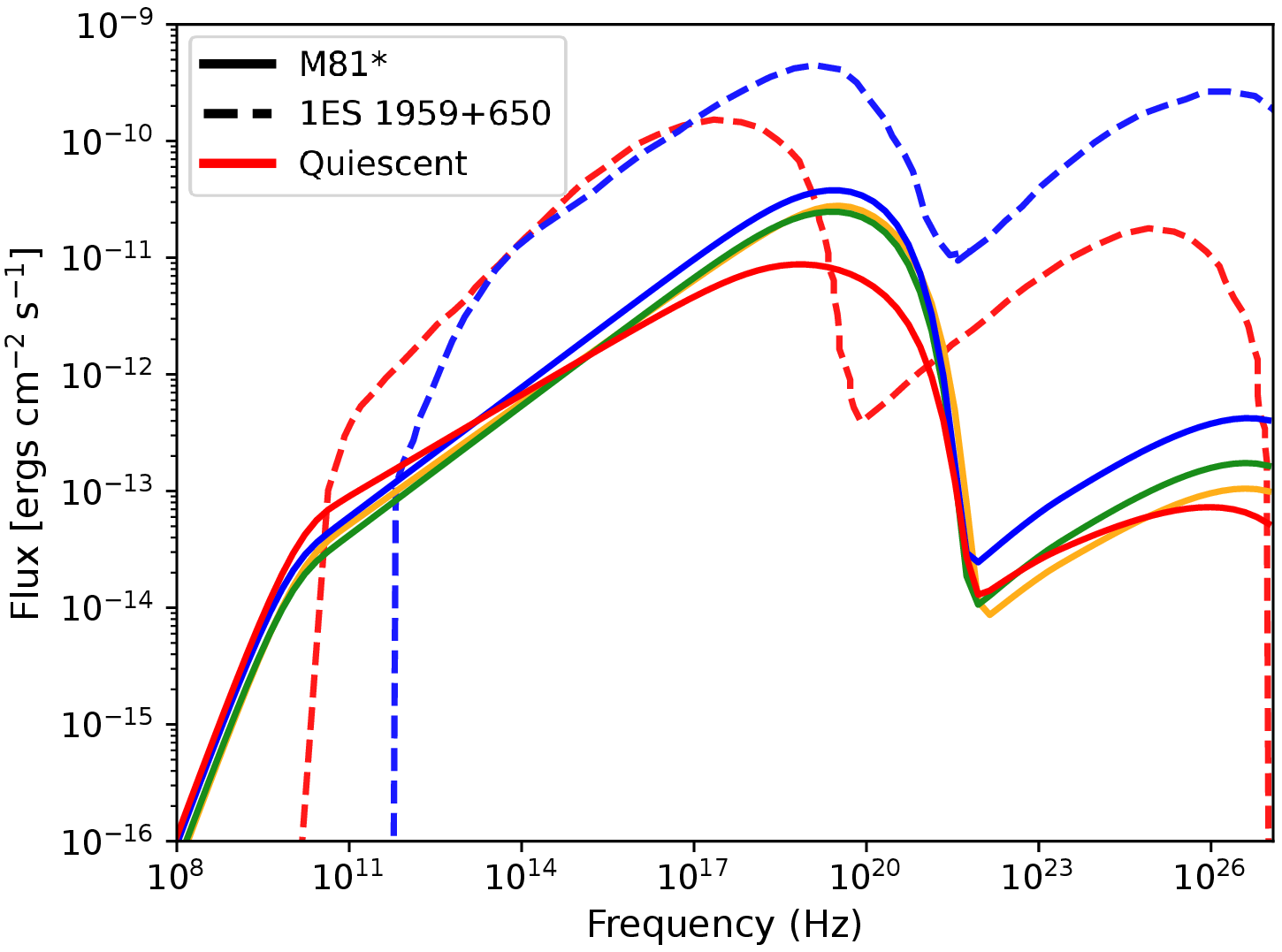}
    \caption{SED of blazar 1ES 1959+650 obtained using a one-zone SSC model during quiescent state (red dashed line; reproduced from \citealt{tag_qui08}) and flaring state (blue dashed line; reproduced from \citealt{1es_magic20}) compared with that of M81* during quiescent state (red solid line) and flaring states (blue, green, and yellow solid lines).}
    \label{comparison}
\end{figure*}

Figure \ref{par} shows how the values of two of the observables (X-ray photon index, ($\Gamma_x$) and flux) and our model parameters (electron spectral index at injection, ($\Gamma_e$), the energy density in injected electrons in the comoving frame ($U_{e,inj}'$), magnetic field ($B$), and jet power in injected electrons, ($P_{e,inj}$) change in different states of M81* that we have considered in our study.

 A hard-injected electron spectrum is required to explain the hardness in the X-ray spectrum, which is observed during Flare B and Flare C. At the other epochs too, injected electron spectrum evolution is found to be in accordance with the evolution of the X-ray spectrum. 
 The jet remains dominated by particles during all the states, and equipartition in energy is not maintained.
 
 \par
 During the flares, the X-ray flux increases but the shape of the SED does not change significantly from the quiescent state. The synchrotron peak flux increases, and the peak shifts slightly towards higher energy in the flaring states. The injected jet power in electrons is found to increase with increasing X-ray flux during flares. 
 While the X-ray flux is lower at Epoch 4 than the quiescent state, the jet power is the highest at this epoch. This could be attributed to the increase in radio emissions.
 
 \par
 LLAGNs are interesting sites to study cosmic ray acceleration, multi-wavelength flares, the time evolution of particle spectrum, and jet parameters. We conclude from the SED modeling of the different states and epochs of M81* that the synchrotron emission from the one-zone model can explain the radio to X-ray emission from M81*. Detection of gamma rays can help to constrain the model better. Low values of Lorentz bulk factor, magnetic field, and jet power in electrons compared to those of the high luminosity AGNs are common features of LLAGNs.
With more observations on the spectral and temporal evolution of LLAGNs, it would be possible to confirm whether the jet activities of LLAGNs and high-luminosity AGNs are similar.



\begin{table*}
    
    \resizebox{\textwidth}{!}{\begin{tabular}{ l l l l l l l l }
    \hline
    Parameter & Quiescent State & Flare A & Epoch 2 & Epoch 3 & Epoch 4 & Flare B & Flare C \\
    \hline
   
    Spectral index & 2.42 & 2.3 & 2.3 & 2.3  & 2.57  & 2.26 & 2.26\\
    $\Gamma_{min}$& 400 & 400 & 400  & 400 & 580 & 400 & 400\\
    $\Gamma_{max}$& 1.5$\times10^{8}$ & 1.5$\times 10^8$ & 1.5$\times 10^8$ & 1.5$\times10^8$ & 1.5 $\times$10$^8$ & 1.5$\times10^8$ & 1.5$\times10^{8}$ \\
    R [cm]& 9.$\times10^{16}$ & 9.$\times10^{16}$ & 9.$\times10^{16}$ & 9.0$\times10^{16}$  & 9.0$\times$10$^{16}$& 9.0$\times10^{16}$ & 9.0$\times10^{16}$\\
    B [Gauss]& 0.004 & 0.004 & 0.004 & 0.004  & 0.004 & 0.003  & 0.003 \\ 
    Doppler factor& 4.1 & 4.1 & 4.1 &  4.1 & 4.1 & 4.1 & 4.1\\
    Bulk $\Gamma$ & 3.7 & 3.7 & 3.7 & 3.7  & 3.7 & 3.7 & 3.7\\
    U$_{e,inj}'$[erg cm$^{-3}$]& 1.19$\times10^{-11}$& 1.57$\times10^{-11}$ & 4.20 $\times10^{-11}$ & 7.70$\times10^{-11}$  & 1.17$\times$10$^{-10}$& 1.21$\times10^{-11}$ & 1.98$\times10^{-11}$\\
    P$_{e,inj}$[erg s$^{-1}$]& 1.25$\times10^{35}$ & 1.64$\times10^{35}$ & 4.39 $\times10^{35}$ & 8.06$\times10^{35}$  & 1.64$\times$10$^{36}$& 1.27$\times10^{35}$ & 2.07$\times10^{35}$\\
    U$_B'$/U$_e'$ & 0.017 & 0.027 & 0.020 & 0.017  & 0.014 & 0.008 & 0.007\\
    \hline
    \end{tabular}}
  \caption{Parameters obtained from the Modeling of the SEDs of M81* in Different States. Note, the U$_{e,inj}'$ and U$_B'$ are the energy densities in electrons and magnetic field in the jet frame, P$_{e,inj}$ is the jet power in injected electrons. The escape time for each case is taken to be R/c.}  
  \label{table_par}
\end{table*}


\section*{Data Availability}
All the data used for this study is publicly available online. The observational X-ray data for \swift~ and \textit{NuSTAR} are available at HEASARC archive\footnote{\url{https://heasarc.gsfc.nasa.gov/cgi-bin/W3Browse/w3browse.pl}}. This work has also made use of public \textit{Fermi-LAT} data obtained from Fermi Science Support Center (FSSC), provided by NASA Goddard Space Flight Center.

\section*{Software}
SAS (v18.0; \cite{gabriel}), HEAsoft (v6.26.1; \url{https://heasarc.gsfc.nasa.gov/docs/software/heasoft/}), XSPEC (v12.10.0f; \cite{arnaud}), fermipy (v1.0.0;\citealt{fermipy}), GAMERA (\citealt{hahn}).
\section*{Acknowledgements}
We thank the referee for their constructive comments, which have helped improve the manuscript. G.T. thanks Jean-Philippe Lenain for the helpful discussion on Fermi-LAT analysis for M81*. 
\bibliography{m81}
\bibliographystyle{aasjournal}






\end{document}